\documentclass[twocolumn]{article}

\RequirePackage[utf8]{inputenc}
\RequirePackage[english]{babel}
\RequirePackage{amsmath,amsfonts,amssymb}
\RequirePackage{mathpazo} 
\RequirePackage[scaled]{helvet}
\RequirePackage[T1]{fontenc}
\RequirePackage{url}
\RequirePackage[colorlinks=true, allcolors=blue]{hyperref}

\newcommand{\titlefont}{\normalfont\bfseries\fontsize{22}{25}\selectfont}

\RequirePackage{authblk}
\usepackage[amssymb]{SIunits}
\usepackage{graphicx}
\usepackage{booktabs}
\usepackage{xcolor,colortbl}
\definecolor{lightblue}{cmyk}{0.1,0,0,0}
\definecolor{lightgreen}{cmyk}{0.1,0,0.1,0}
\definecolor{lightred}{cmyk}{0,0.1,0,0}

\RequirePackage{graphicx,xcolor}
\RequirePackage{colortbl}
\RequirePackage{booktabs}
\RequirePackage{algorithm}
\RequirePackage[noend]{algpseudocode}
\RequirePackage{changepage}
\RequirePackage[left=48pt,%
                right=42pt,%
                top=46pt,%
                bottom=60pt,%
                headheight=15pt,%
                headsep=10pt,%
                letterpaper]{geometry}%
\RequirePackage[labelfont={bf,sf},%
                labelsep=period,%
                figurename=Fig.]{caption}
\definecolor{color0}{RGB}{0,0,0} 
\definecolor{color1}{RGB}{59,90,198} 
\definecolor{color2}{RGB}{16,131,16} 
   
\RequirePackage[numbers,sort&compress]{natbib}
\RequirePackage[resetlabels]{multibib}
\newcites{fullrefs}{Full References}  
\bibliographystylefullrefs{osajnl}    

\RequirePackage{letltxmacro}
\RequirePackage{xparse}
\LetLtxMacro\oldcite\cite
\RenewDocumentCommand{\cite}{O{} O{} m}{\oldcite[#1][#2]{#3}\nocitefullrefs{#3}}
        
\arrayrulecolor{color0} 
\RequirePackage[explicit]{titlesec}
\renewcommand{\thesubsection}{\Alph{subsection}}

\titleformat{\section}
  {\large\bfseries}
  {\thesection.}
  {0.5em}
  {\MakeUppercase{#1}}
  []
\titleformat{name=\section,numberless}
  {\large\sffamily\bfseries}
  {}
  {0em}
  {\MakeUppercase{#1}}
  []  
\titleformat{\subsection}
  {\sffamily\bfseries}
  {\thesubsection.}
  {0.5em}
  {#1}
  []
\titleformat{\subsubsection}
  {\sffamily\small\bfseries\itshape}
  {\thesubsubsection.}
  {0.5em}
  {#1}
  []    
\titleformat{\paragraph}[runin]
  {\sffamily\small\bfseries}
  {}
  {0em}
  {#1} 
\titlespacing*{\section}{0pc}{3ex \@plus4pt \@minus3pt}{5pt}
\titlespacing*{\subsection}{0pc}{2.5ex \@plus3pt \@minus2pt}{2pt}
\titlespacing*{\subsubsection}{0pc}{2ex \@plus2.5pt \@minus1.5pt}{2pt}
\titlespacing*{\paragraph}{0pc}{1.5ex \@plus2pt \@minus1pt}{12pt}

\renewcommand{\@maketitle}{%
{%
\vskip68pt%
{\raggedright \titlefont \@title\par}%
\vskip11pt
{\raggedright \@author\par}
\vskip11pt%
{
\vskip25pt%
}%
}%
}%

\RequirePackage{enumitem} 
\usepackage{lipsum} 

\title{Efficient fiber-coupled single-photon source based on quantum dots in a photonic-crystal waveguide}

\author[1]{Rapha\"el S. Daveau}
\author[2,3]{Krishna C. Balram}
\author[1]{Tommaso Pregnolato}
\author[2,3]{Jin Liu}
\author[4]{Eun H. Lee}
\author[4]{Jin D. Song}
\author[5]{Varun Verma}
\author[5]{Richard Mirin}
\author[5]{Sae Woo Nam}
\author[1]{Leonardo Midolo}
\author[1]{S\o ren Stobbe}
\author[2,*]{Kartik Srinivasan}
\author[1,$\dagger$]{Peter Lodahl}

\affil[1]{Niels Bohr Institute, University of Copenhagen, Blegdamsvej 17, DK-2100 Copenhagen, Denmark}
\affil[2]{Center for Nanoscale Science and Technology, National Institute of Standards and Technology, Gaithersburg, MD 20899, USA}
\affil[3]{Maryland NanoCenter, University of Maryland, College Park, MD 20742, USA}
\affil[4]{Center for Opto-Electronic Convergence Systems, Korea Institute of Science and Technology, Seoul 136-791, Korea}
\affil[5]{National Institute of Standards and Technology, Boulder, CO 80305, USA}
\affil[*]{Corresponding author: kartik.srinivasan@nist.gov}
\affil[$\dagger$]{Corresponding author: lodahl@nbi.ku.dk}

\begin{document}
\maketitle

\textbf{Many photonic quantum information processing applications would benefit from a high brightness, fiber-coupled source of triggered single photons. Here, we present a fiber-coupled photonic-crystal waveguide single-photon source relying on evanescent coupling of the light field from a tapered out-coupler to an optical fiber. A two-step approach is taken where the performance of the tapered out-coupler is recorded first on an independent device containing an on-chip reflector.  Reflection measurements establish that the chip-to-fiber coupling efficiency exceeds 80~\%. The detailed characterization of a high-efficiency photonic-crystal waveguide extended with a tapered out-coupling section is then performed. The corresponding overall single-photon source efficiency is 10.9~\% $\pm$ 2.3~\%, which quantifies the success probability to prepare an exciton in the quantum dot, couple it out as a photon in the waveguide, and subsequently transfer it to the fiber. The applied out-coupling method is robust, stable over time, and broadband over several tens of nanometers, which makes it a highly promising pathway to increase the efficiency and reliability of planar chip-based single-photon sources.}

\begin{figure*}[t!]
  \centering
  \includegraphics[width=\linewidth]{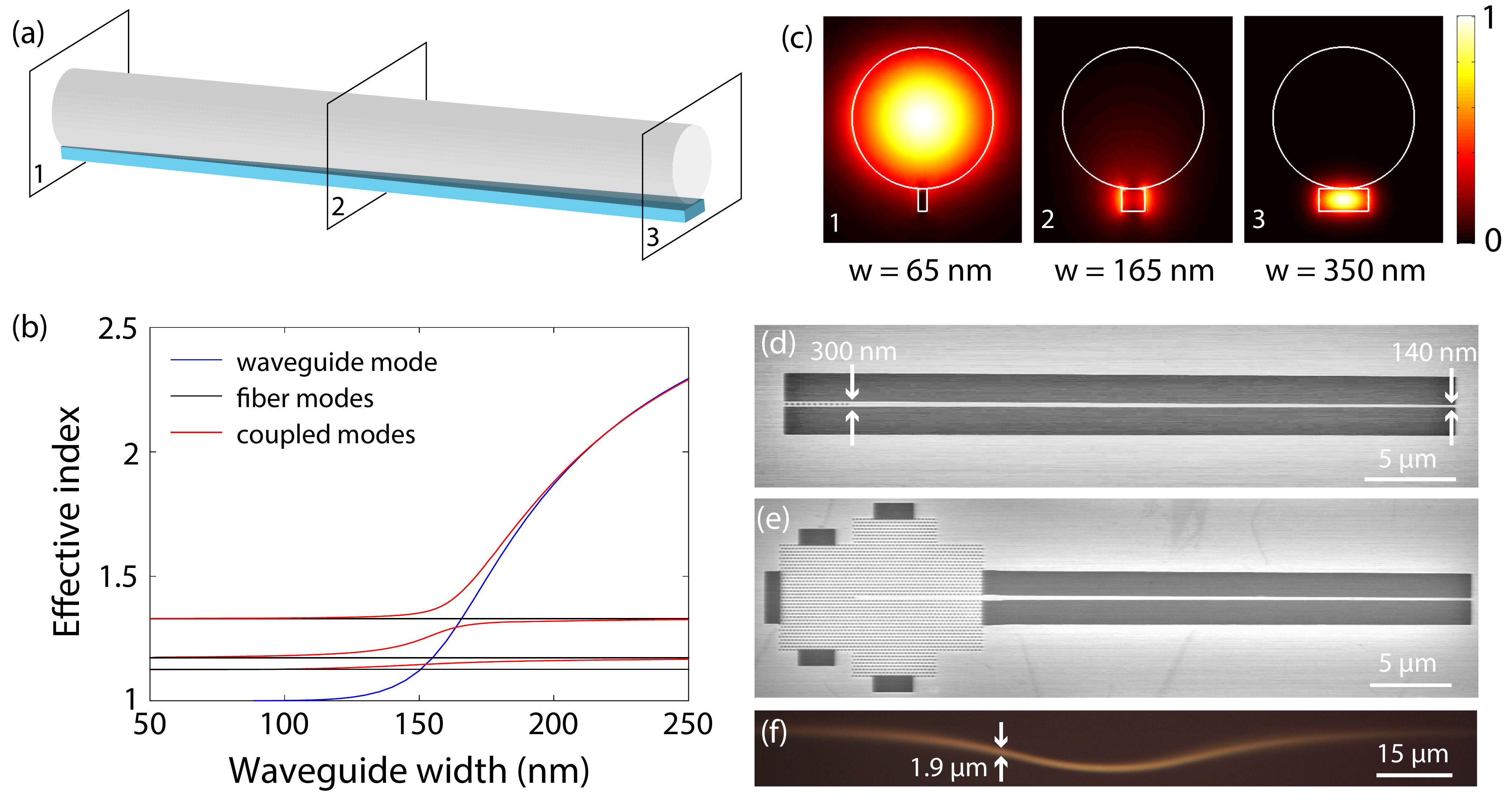}
  \caption{Evanescent coupler design and devices. (a) Sketch of the coupled waveguide system. The microfiber (gray cylinder) with a diameter of 1~$\micro\meter$ is brought in contact with the tapered GaAs waveguide (blue taper) of thickness $t=160$~nm. The waveguide is simulated with a width varying from 50~nm to 350~nm. (b) Effective index of the bare tapered waveguide (blue), bare microfiber (black), and coupled waveguide-fiber system (red), as obtained from finite-element simulations. (c) Normalized total electric field amplitude $|\mathbf{E}|$ plotted in the cross section plane of the coupled waveguide system (as marked in (a)) for three different waveguide widths $w$. (d) Scanning-electron micrograph of the 30~$\micro\meter$ long tapered nanobeam waveguide (right portion of image) terminated with a 1D photonic-crystal mirror (left portion of image). (e) Scanning-electron micrograph of the photonic-crystal waveguide (left portion of image) connected to the 30~$\micro\meter$ long tapered waveguide (right portion of image) for off-chip coupling. (f) Optical microscope image of the dimpled microfiber, with actual diameter of 1.9~$\micro\meter$ $\pm$ 0.2~$\micro\meter$.}
  \label{fig:fig1}
\end{figure*}

\section*{Introduction}
Single photons are very robust carriers of quantum information that can be used for quantum simulations as well as other applications in quantum technology \cite{Obrien2009,Aspuru2012photonic,Walmsley2015}. There is a strong need to develop reliable and practical single-photon sources delivering on demand, highly pure, and coherent single photons. Recent works have demonstrated tremendous progress on all of these features \cite{Lodahl2015interfacing}, in particular in micropillar devices relying on the vertical out-coupling of photons from quantum dots \cite{Somaschi2016,Ding2016,Loredo2016}, but the low out-coupling efficiency of photons into a fiber remains the main bottleneck. A different approach is based on the \emph{planar technology} of nanophotonic waveguides, which present two main advantages: 1) Emitter-waveguide coupling efficiencies ($beta$-factor) exceeding 98~\% have been demonstrated thus constituting a deterministic photon-emitter interface \cite{Arcari2014} so that such a platform can be used as a building block for highly efficient single-photon sources. 2) Planar waveguide sources can be readily integrated in complex photonic circuits for fully on-chip photonic quantum-information processing.

To this date, it remains a challenge to efficiently out-couple photons from such an inherently efficient single-photon source, because of the high refractive index of the materials and the size mismatch between on-chip nanophotonic waveguides and optical fibers. Previous work on chip-to-fiber interfaces in silicon-based photonics employed either optimized gratings \cite{Zaoui2014} or end-fire coupling from a tapered waveguide to a cleaved fiber \cite{Cohen2013}. The most efficient technique relies on evanescent coupling from a microfiber to an on-chip waveguide \cite{Groblacher2013,Barclay2004} with a record chip-to-fiber coupling efficiency of 97~\% \cite{Tiecke2015}. Use of evanescent coupling to collect single photons from QDs~\cite{Davancco2011} or nitrogen-vacancy centers in diamond \cite{Patel2016efficient}, has been recently studied, with single-photon source efficiencies into the fiber of 6~\% and 4.2~\%, respectively.

In the present work, single photons efficiently generated from a quantum dot coupled to a unidirectional gallium arsenide photonic-crystal waveguide are evanescently coupled to a microfiber on a separate tapered waveguide out-coupler. The tapering of the waveguide allows for an irreversible power transfer from the waveguide to the fiber, rather than the oscillatory power transfer (along the propagation direction) that would occur for a fixed waveguide width~\cite{Davancco2011}. The unprecedented internal efficiency of the source in conjunction with the ability to produce highly indistinguishable photons \cite{Kirsanske2016} opens very promising prospects for implementing optimal planar photonic-crystal waveguide sources, crucial in applications such as boson sampling \cite{He2016scalable,Loredo2016bosonsampling}.

\section*{Design of the chip-to-fiber out-coupling taper}

Quantum dots coupled to photonic-crystal waveguides (PCWG) make an intrinsically deterministic single-photon source, because the emitter-waveguide coupling is in theory  $\beta\to$ 1 \cite{rao2007single}, which has been exemplified in experiments \cite{Arcari2014}. The remaining challenge is to couple the collected photons out of the waveguide to an optical fiber with high efficiency. To this end, an efficient out-coupler is designed as illustrated in Fig.~\ref{fig:fig1}(a), which is adapted from Gr\"{o}blacher \textit{et al.} \cite{Groblacher2013} to the case of GaAs at a wavelength of 940~nm. The operational principle relies on evanescent coupling from a tapered waveguide to a microfiber after being brought into contact with each other.

The effective indices of the taper and the fiber as well as the coupled fiber-taper system are calculated with finite element method and displayed in Fig.~\ref{fig:fig1}(b). The refractive indices used for the simulation are 3.46 for the GaAs waveguide and 1.45 for the silica fiber.  The fiber (black lines) has a constant diameter, assumed to be 1~$\micro\meter$, along the full interaction region, while the fundamental transverse-electric (TE) mode of the 160~nm thick GaAs waveguide (blue line) is a function of its width $w$. When the two waveguides are brought in contact, the coupled system supports a series of mixed supermodes (red lines) due to the interaction between the different bands. For $w> 200$~nm, the waveguide mode has an effective index above 2 and the fiber has barely any effect on the mode, which remains mostly in the waveguide. For  200~nm $<w<$ 140~nm, the effective index of the waveguide mode approaches that of the fiber, i.e., the mode gets transferred to the fiber. For $w<140$~nm, the effective index of the waveguide mode goes to 1 and the mode is mostly confined in the fiber. For example, for a tapering width of $w=140$~nm, a 95~\% mode overlap between the bare fiber mode and the coupled waveguide mode is calculated by evaluating the mode overlap integral. Figure~\ref{fig:fig1}(c) plots the normalized electric field amplitude in the cross section of the coupled waveguide system and illustrates how the mode is transferred from the waveguide to the fiber as the waveguide width increases.

The design of the waveguide taper is made adiabatic, i.e., the width varies slowly along the propagation direction to prevent the fundamental mode from coupling to higher order supermodes, according to the condition $\mathrm{d}n_\mathrm{WG}/\mathrm{d}y \ll (2\pi/\lambda_0)|n_\mathrm{eff,1} - n_\mathrm{eff,2}|^2$ \cite{Groblacher2013}, where $ n_\mathrm{WG} $ is the effective index of the bare waveguide mode, $ n_\mathrm{eff,i} $ is the effective index of the i$^\mathrm{th}$ coupled waveguide mode displayed in Fig.~\ref{fig:fig1}(b), and $\lambda_0=940$~nm is the free-space wavelength. The tapers are designed to have a length $L = 30$~$\micro\meter$.

Two types of samples are fabricated, each containing self-assembled InAs quantum dots (QDs) grown by molecular beam epitaxy, with a QD density of 100~$\micro\meter^{-2}$ and emitting in the 920~nm $\pm$ 25~nm range. They are embedded in a 160~nm thick GaAs membrane, where waveguides and nanostructures are fabricated by means of electron-beam lithography and dry and wet etching \cite{Midolo2015soft}.

The first type of sample, cf. Fig.~\ref{fig:fig1}(d), serves the purpose of  testing the quality of the taper and the out-coupling to the fiber. It contains a tapered nanobeam waveguide (NWG) ending in a 1D-photonic crystal reflector whose photonic bandgap spans from 870~nm to 1060~nm. The position and size of the first two holes of the reflector are modified adiabatically to prevent out-of-plane scattering of the incoming wave \cite{Sauvan2005modal}. The waveguide is tapered down from a width of 300~nm at the reflector end to a final width of 140~nm over a length of 30~$\micro\meter$.

The second sample is shown in Fig.~\ref{fig:fig1}(e) and features a unidirectional PCWG terminated on one side by a photonic-crystal section operating as a mirror. The PCWG has a 5~$\micro\meter$ long slow-light section (measured band edge at 915~nm) to enhance the light-matter coupling in order to reach a near-unity $\beta$-factor \cite{Arcari2014}. The PCWG is coupled to a tapered NWG of a similar type as the one studied above, however tapered down to a final width of 160~nm. The transition from the slow-light section of the PCWG to the tapered NWG is naturally prone to reflection loss due to impedance mismatch between the two optical modes. In the device, this loss is minimized by engineering a modified PCWG section supporting light with a larger group velocity (fast-light section) \cite{Hugonin2007coupling}, whose band edge is at 980~nm. Finite-element method simulations yield a transmission efficiency at the interface between PCWG and tapered NWG of 88~\% at 940~nm.

Finally, the out-coupling single-mode fiber is tapered by a heat-and-pull technique and imprinted with a dimple \cite{Michael2007optical}, as shown in Fig.~\ref{fig:fig1}(f), to avoid touching the chip with the entire fiber, thus enabling the interaction with single nanophotonic waveguides \cite{Hauer2014}. The diameter of the fiber at the position of the dimple is 1.9~$\micro\meter$ $\pm$ 0.2~$\micro\meter$, measured with a calibrated microscopy setup, where the uncertainty is due to inaccuracy in edge detection and is a one standard deviation value. The larger diameter of the fabricated tapered fiber compared to the designed value increases its effective index and therefore the cross-over point in Fig.~\ref{fig:fig1}(b). However, this only slightly shifts the optimal working wavelength of the coupler while the efficiency should remain unchanged or increased. The tapered fiber is then clamped into a U-shape, mounted and fed outside of the cryostat chamber by using a teflon feedthrough \cite{Srinivasan2007optical}. The overall measured transmission of the tapered fiber (including input and output splices and connectors) is 67.4~\% $\pm$ 1.5~\%, where the uncertainty is due to fluctuations in the detected power levels and represents a one standard deviation value. The non-ideal transmission level is due to the taper and dimple sections, as well as dust residues caught by the tapered part of the fiber while mounted in the cryostat. For the out-coupling from the device, light propagates only through half of the length of the fiber, i.e., the relevant fiber transmission assuming the losses are symmetric is $\sqrt{0.674} = 82.1$~\% $\pm$ 1.8~\%, which is measured at 930~nm. All experiments are carried out inside a closed-cycle cryostat at 4.6 K.

\section*{Broadband characterization of the chip-to-fiber coupling efficiency}

\begin{figure}[t!]
  \centering
  \includegraphics[width=\linewidth]{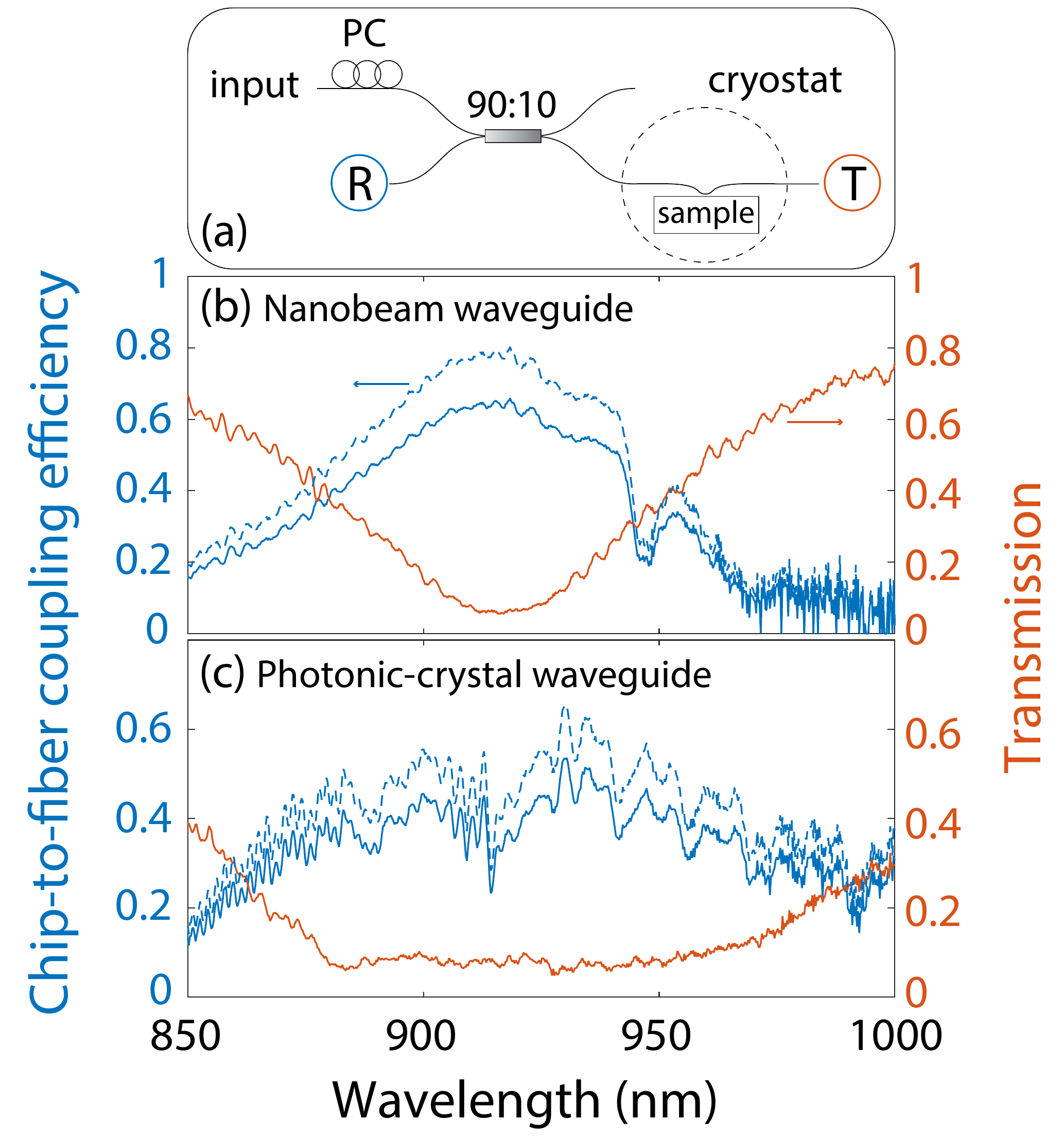}
  \caption{Passive characterization of the chip-to-fiber coupling through reflection and transmission measurements. (a) Sketch of the optical setup with polarization control (PC) and 90:10 fiber beam splitter. (b) Chip-to-fiber coupling efficiency $ \eta_\mathrm{CF}$ for the NWG (solid blue line) obtained from reflection measurement and extracted as explained in the text. The dashed blue line data are obtained after correcting for the measured propagation losses in the tapered fiber. The red curve shows the transmission through the fiber when coupled to the device relative to the case of an uncoupled fiber, i.e., it also corrects for propagation losses in the tapered fiber. (c) Similar data for the device containing the PCWG.}
  \label{fig:fig2}
\end{figure}

The chip-to-fiber coupling efficiency is an important figure-of-merit characterizing a fiber-coupled single-photon source. It is first investigated with the NWG sample (Fig.~\ref{fig:fig1}(d)), which was specifically designed for this purpose. We monitor the reflection and transmission through the fiber connected to the device by employing a pulsed supercontinuum white-light source filtered with a 830~nm long-pass filter and 1060~nm short-pass filter in order to probe a wide wavelength range around the designed wavelength of 940~nm. The setup used for the characterization is illustrated in Fig.~\ref{fig:fig2}(a) and consists of a polarization control unit and a 90:10 fiber beam splitter. The spectrum is recorded with a fiber-coupled spectrometer either at the input of the probe fiber (reference), output of the probe fiber (transmission) or the counter-propagating output of the exit port of the beam splitter (reflection). The polarization of the laser is controlled in order to launch a TE-like mode from the fiber into the tapered waveguide.

The chip-to-fiber coupling efficiency $ \eta_\mathrm{CF} = \sqrt{P_R/P_I\eta_\mathrm{FBS}} $ is obtained by recording the reflected power $ P_R $ and the input power $ P_I $. The transmission efficiency of the 90 port of the fiber beam splitter $\eta_\mathrm{FBS}$ has also been measured with the same broadband source to account for its spectral dependence. The coupling efficiency is plotted (solid blue line) for the NWG in Fig.~\ref{fig:fig2}(b), where we observe  $ \eta_\mathrm{CF} >65$~\% around 920~nm. Correcting for the insertion and propagation loss in the tapered fiber (dashed blue line) leads to $ \eta_\mathrm{CF} >80$~\%. We emphasize that this constitutes a conservative lower bound on the actual coupling efficiency, since the recorded spectrum reflects on-chip at the photonic-crystal mirror, which may have a reflectivity below 1 due to fabrication imperfections. Excitingly, the chip-to-fiber coupling efficiency of $>80$~\%, obtained from reflection measurements on samples in a GaAs platform containing active QD emitters, is approaching a level that has so far only been reported for passive samples based on silicon \cite{Barclay2004,Groblacher2013,Tiecke2015}.

Another indication of the very high coupling efficiency is obtained by transmission measurements through the fiber, by comparing the transmitted power when the fiber is coupled to the chip device to the power when it is not coupled, cf. Fig.~\ref{fig:fig2}(b). We observe that the transmission drops below 10~\% when the fiber is connected to the device. The fact that the transmission and reflection data have the same spectral dependence is a sign that the coupling to the device is the dominating extinction channel. However, we notice that the transmission values do not return fully to unity away from the coupler resonance which is either due to the very broadband feature of the coupler not grasped by the wavelength range investigated here, or due to a broadband insertion loss when coupling the fiber to the device. The broadband insertion loss may be due to the sharp change in refractive index created at the interface where the taper tip is attached to the suspended membrane. The mode launched from the fiber into the device can undergo partial direct back-reflection or scattering at such an interface.

We subsequently reproduce the same type of measurement with the sample containing the PCWG shown in Fig.~\ref{fig:fig1}(e) and the results are plotted in Fig.~\ref{fig:fig2}(c). The measurements exhibit an extinction of $ \approx 90$~\% of the transmitted power, which is fully consistent with the measurements on the NWG since the out-coupling structure is similar. From reflection measurements we observe $ \eta_\mathrm{CF} >60$~\% around 930~nm after correcting for the insertion and propagation losses in the tapered fiber. The lower reflection compared to the NWG device stems from the fact that the PCWG device contains additional interfaces (tapered waveguide to fast-light PCWG and fast-light PCWG to slow-light PCWG) and therefore losses. Consequently, the recorded light propagates from the tapered waveguide and into the PCWG before being reflected and subsequently transferred back to the tapered waveguide from the PCWG. The simulated one-way transmission efficiency at the interface between PCWG and tapered waveguide is 88~\% at 940~nm, which explains the overall lower reflected power with this device compared to the NWG.

We find that both devices have a maximum coupling efficiency around 920~nm, but with varying bandwidths. We anticipate that the bandwidth varies with the position of the 15~$\micro\meter$ long fiber dimple over the extent of the  30~$\micro\meter$ long waveguide taper. Also it should be mentioned that the NWG device was tapered to a smaller width of 140~nm as opposed to 160~nm for the PCWG.

\section*{Single-photon source efficiency}

\begin{figure*}[t]
  \centering
  \includegraphics[width=\linewidth]{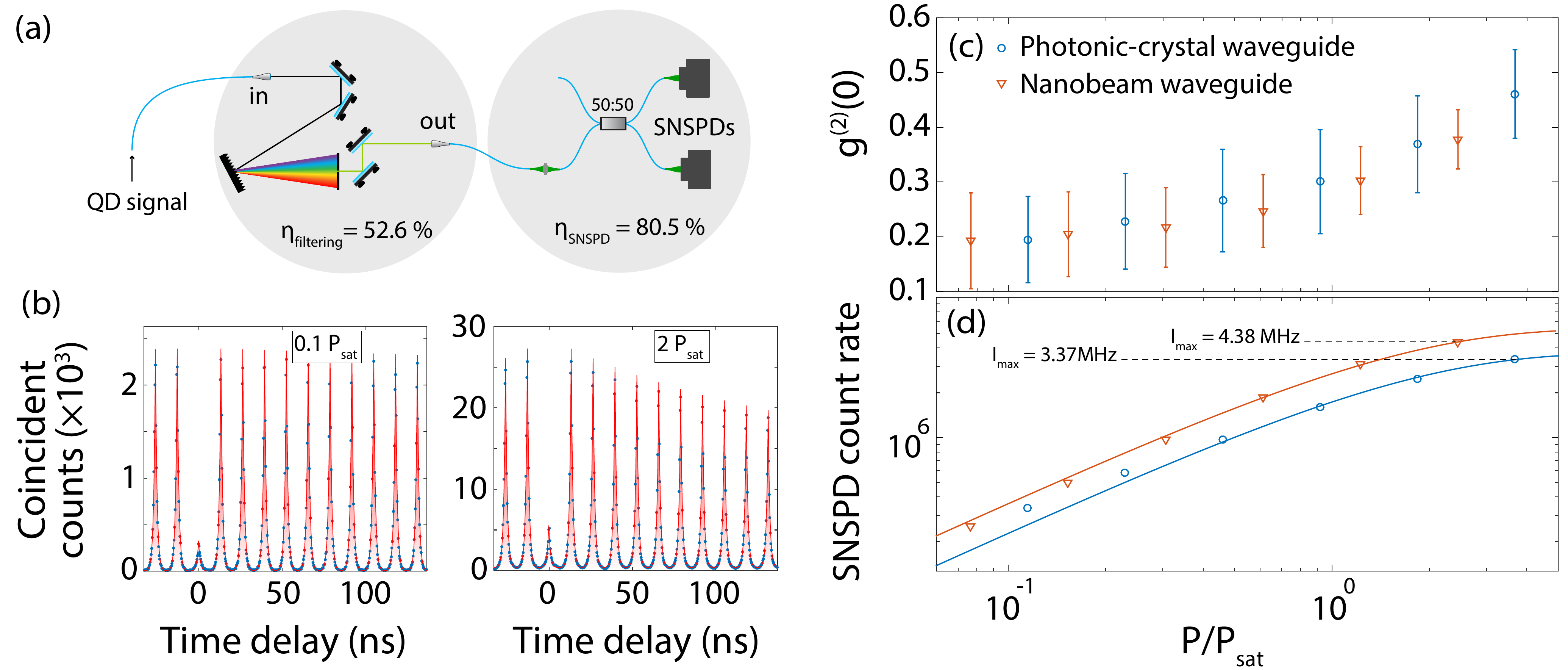}
  \caption{Characterization of QDs in a PCWG and a NWG. (a) Detection setup: the QD emission collected from the device is filtered and sent to superconducting nanowire single-photon detectors (SNSPDs). (b) Auto-correlation measurement at 0.1 and 2 $P_\mathrm{sat}$ for a QD coupled to a PCWG. The data are fitted (red curve) to extract the value of $g^{(2)}(0)$. (c) Auto-correlation function at zero time delay as a function of pumping power for a QD coupled to a NWG (red triangle) and PCWG (blue circle). Error bars are 95~\% confidence interval from the fits. (d) Raw SNSPD count rates for the brightest QDs in the two types of waveguides versus excitation power. The saturation behavior of the QDs is confirmed by fitting the data with $I = I_\mathrm{max}(1-\exp(-P/P_\mathrm{sat})) $ (solid lines), which allows to extract the maximum achievable count rate $I_\mathrm{max}$ and the saturation pump power $P_\mathrm{sat}$.}
  \label{fig:fig3}
\end{figure*}

Several parameters determine the overall efficiency of a single-photon source including the probability to create a single excitation in the quantum emitter upon optical pumping, to collect the emitted photon with the photonic nanostructure, and to subsequently route the photon into the usable mode of an optical fiber. The resulting total probability that the optical excitation of the QD leads to a propagating photon in the fiber is denoted \emph{the source efficiency}, which is the figure-of-merit for applications of the single-photon source. To extract the source efficiency, we study single QDs embedded in the nanostructures, while exciting and collecting through the fiber. The chip-to-fiber coupling is optimized by maximizing the recorded emission intensity of the QD ensemble excited with a continuous-wave laser at 860~nm while repeatedly landing the fiber on the taper and slightly adjusting its position. Subsequently single QD lines are investigated by redshifting the laser wavelength in order to excite through the $p$-shell transition of one of the brightest QD lines on the sample. The laser is subsequently operated in a pulsed mode (3~ps pulse) with a 76~MHz repetition rate and the saturation curve of the QDs is recorded by measuring the emission intensity of the QD line versus excitation power on a spectrometer. Examples of saturation curves are displayed in Fig.~\ref{fig:fig3}(d) and by modeling the data we obtain the characteristic saturation power $P_\mathrm{sat} $ of the QD. Furthermore, we record the single-photon purity of the source by performing a second-order correlation measurement of the QD signal. The QD line is filtered using a grating-pinhole setup (see Fig.~\ref{fig:fig3}(a)) with a bandwidth of 0.5~nm and a transmission efficiency of 52.6~\% $\pm$ 4.2~\%, which is calibrated using a single-frequency laser tuned to the QD emission wavelength. We note that the QD emission also features phonon sidebands that typically amount to the level of 10~\% to 20~\% of the emission in the zero-phonon line at a temperature of 4~K \cite{Lodahl2015interfacing}. These phonon sidebands will be removed by the implemented filter setup implying that the actual transmission efficiency is correspondingly lower. The signal is then sent into a fiber-based 50:50 beam splitter connected to two superconducting nanowire single-photon detectors (SNSPDs) \cite{Natarajan2012superconducting,Marsili2013} with a measured detection efficiency of 80.5~\% $\pm$ 4.7~\% at 930~nm. In order to remain within the linear regime, we place a neutral density filter (8.42$\times$ attenuation) before the detectors.

We investigate the brightest QD coupled to the PCWG, which emits at 931~nm and is excited through a resonance of the QD at 905~nm. The second-order correlation function for two different pumping powers is shown in Fig.~\ref{fig:fig3}(b) and is fitted with a sum of exponentially decaying peaks multiplied by slowly decaying exponential functions to account for blinking of the QD transition. Blinking effectively reduces the source efficiency, since the QD switches between bright and dark emitter states, which is commonly observed for quasi-resonant excitation of self-assembled QDs \cite{Santori2001}. We generally observe an increase in blinking rate with pumping power (compare the two traces of Fig.~\ref{fig:fig3}(b)), while the magnitude of the blinking is found to vary for different QDs. In the presence of blinking, the purity of the single-photon pulses, which is gauged by the value of  $g^{(2)}(0)$, can be recorded as the area of the suppressed peak found at zero time delay relative to the average area of the peaks at long times where effects of blinking are negligible \cite{Santori2004}. The blinking amplitude, which assesses the preparation efficiency of the source, is extracted from the fit as the peak amplitude at long time delay over the peak amplitude at zero time delay \cite{Santori2001}. Figure~\ref{fig:fig3}(c) plots $g^{(2)}(0)$ as a function of pumping power, where the error bars are 95~\% confidence interval from the fits. We obtain $g^{(2)}(0)= 0.20 \pm 0.08$ at $0.1 P_\mathrm{sat}$, which quantifies the good quality of the single-photon emission. Despite the quasi-resonant excitation scheme, exciting through the fiber excites multiple QDs in the waveguide (PCWG + tapered waveguide section), hence the non-zero $g^{(2)}(0)$. However, the purity could straightforwardly be improved further by reducing the density of QDs on the structure \cite{Somaschi2016} or by using a narrower band spectral filtering of the QD line.

By recording the decay rate of the QD in the PCWG, we can determine the single-photon coupling efficiency to the waveguide mode (the $\beta$-factor) by employing the method of Ref.~\cite{Arcari2014}. The decay curve is recorded at $0.2 P_\mathrm{sat}$ and exhibits a single-exponential decay, which is expected when exciting the QD quasi-resonantly because bright excitons are predominantly prepared due to the spin-conserving cascade to the ground-state exciton \cite{Tighineanu2016single}. The corresponding decay rate is 1.13 ns$^{-1}$, from which $\beta = 0.91$ is extracted. We mention that the potential minor influences of non-radiative processes are not included in this analysis, which for neutral excitons can be accounted for by analyzing their coupling to dark excitons \cite{Tighineanu2013decay}.

To determine the efficiency of the single-photon source, we analyze the recorded SNSPD count rates close to  saturation. First, the single-photon count rate $\Gamma_\mathrm{SP} $ is obtained after correcting for the residual contribution from background emission through $\Gamma_\mathrm{SP} = \Gamma_\mathrm{SNSPD}(1 - g^{(2)}(0))^{1/2}$ by assuming a Poissonian background \cite{Pelton2002efficient}, where $ \Gamma_\mathrm{SNSPD} $ defines the raw SNSPD count rate. As an example, at maximum intensity of the QD, the SNSPD count rate is 3.37~MHz with an associated $g^{(2)}(0)= 0.46$, meaning that the pure single-photon contribution corresponds to a rate of $\Gamma_\mathrm{SP} = 2.43$~MHz. By correcting for the detection efficiency (80.5~\%), spectral filtering efficiency (52.6~\%), beam-splitter efficiency (84.8~\%) and fiber transmission (82.1~\%), we achieve a single-photon rate of coupling from chip to fiber of 8.24~MHz $\pm$ 1.70~MHz. The resulting source efficiency is derived by relating that number to the repetition rate of the excitation laser, and we find an overall source efficiency of 10.9~\% $\pm$ 2.3~\%. The various efficiencies and performance of the single-photon source and detection setup are summed up in Table~\ref{tab:losses}.

\begin{table}[t]
\centering
    \begin{tabular}{p{4.5cm} r}
    \toprule
	  SNSPD count rate  & 3.37~MHz  \\
  Single-photon count rate    & 2.43~MHz $\pm$ 0.43~MHz  \\
 	\midrule
	\rowcolor{lightblue}
  Fiber transmission &  82.1~\% $\pm$ 1.8~\% \\
\rowcolor{lightblue}
  Fiber beam splitter  & 84.8~\% $\pm$ 4.4~\%\\
\rowcolor{lightblue}
  Spectral filtering & 52.6~\% $\pm$ 4.2~\% \\
\rowcolor{lightblue}
  Detector (SNSPD) & 80.5~\% $\pm$ 4.7~\% \\
 	\midrule
\rowcolor{lightgreen}
 Saturation level & 97.5~\%  \\
\rowcolor{lightgreen}
  Preparation ($1-\mathrm{blinking}$) & 55.8~\% $\pm$ 4~\%  \\
  \rowcolor{lightgreen}
  QD-waveguide coupling: $\beta$  & 91~\% $\pm$ 1~\% \\
\midrule
  Single photons in the fiber  & 8.24~MHz $\pm$ 1.7~MHz  \\
  \textbf{Source efficiency}   & $\mathbf{10.9\:\%\pm 2.3\:\%}$  \\
	\bottomrule
  \end{tabular}
\caption{Performance and efficiencies of the single-photon source based on a QD coupled to a PCWG. Blue: off-chip efficiencies for routing, filtering and detecting the single photons. Uncertainties in the off-chip efficiencies are due to fluctuations in measured powers and represent a one standard deviation value. Green: On-chip efficiencies of the single-photon source. Uncertainties in the on-chip efficiencies are 95~\% confidence interval from the fits. The source efficiency (per pulse probability that the source emits a photon in the fiber) is extracted from the single-photon count rate on the SNSPD relative to the repetition rate of the pump laser after correcting for the off-chip efficiencies (blue). }
\label{tab:losses}
\end{table}

We further analyze the recorded source efficiency by relating it to the different on-chip losses measured independently, cf. Table~\ref{tab:losses}. The main limitation of the source comes from the blinking of the QD to off states (44.2~\% at maximum intensity), which by itself diminishes the brightness of the source roughly by half. Considering the measured total on-chip efficiency of single-photon generation (green entries in Table~\ref{tab:losses}: 49.5~\%), off-chip efficiency of routing, filtering and detecting the single photons (blue entries in Table~\ref{tab:losses}: 29.5~\%) and the chip-to-fiber coupling efficiency measured from passive reflection at 931~nm (cf. Fig.~\ref{fig:fig2}(c): 49.6~\%), the expected count rate at the detector is $\Gamma_{\mathrm{SP}} = 5.50$~MHz, which is higher than the experimentally recorded value of 2.43~MHz. This indicates that additional loss processes contribute to the experiment on QDs than accounted for. One possible explanation is the QD fine structure although it was not explicitly tested whether the studied QD line arises from a neutral exciton: a neutral exciton in a QD can be prepared in two (orthogonal) linearly polarized states, which generally couple with very different efficiencies to the PCWG \cite{Lodahl2015interfacing}, and in a quasi-resonant excitation scheme both dipoles may be excited. A non-quasi-resonant excitation of the QD would in fact prepare both bright and dark excitons randomly \cite{Tighineanu2013decay}, which would impact the preparation efficiency significantly, depending on the coupling of the two bright dipoles to the PCWG and the quantum efficiency of the QD \cite{Lodahl2015interfacing}. Furthermore, the filtering of the phonon sidebands by the grating setup will reduce the recorded count rate. Finally, the blinking amplitude extracted from the fit is only a lower estimate because blinking processes that occur faster than the excitation period ($\approx13$~ns) are smeared out because of the pulsed excitation. These effects would have to be studied in continuous wave excitation.

Although not explicitly designed for the purpose of actively collecting single photons, we also check the emission properties of a QD embedded directly in a NWG (cf. Fig.~\ref{fig:fig1}(d)). We find a QD coupled to the NWG emitting at 933~nm, which we excite through a QD resonance at 898~nm. The data are plotted in Figs.~\ref{fig:fig3}(c) and (d). At the highest intensity, we record a maximum SNSPD count rate of 4.38~MHz with corresponding $g^{(2)}(0)= 0.38~\pm0.05$. By performing the same analysis as with the PCWG device, we get a pure single-photon rate on the SNSPD of 3.44~MHz, single-photon pick-up rate directly from chip to fiber of 11.67~MHz and source efficiency of 15.5~\% $\pm$ 2.8~\%. The fiber-coupled brightness demonstrated here is approaching the level of evanescently coupled microfiber-microcavity systems which have source efficiencies in the 20~\% to 25~\% range~\cite{Ates2013improving,Lee2015efficient}, while it readily offers a 2-fold improvement as compared to Ref.~\cite{Davancco2011} using evanescent coupling from a planar waveguide containing QDs to a microfiber. However, in comparison to these previous QD single-photon sources based on evanescent coupling to optical fibers, our approach separates the high $\beta$-factor extraction structure from the fiber out-coupling structure, enabling independent design and optimization of these two critical elements. This also demonstrates that the inherently efficient, robust, and broadband planar single-photon source technology is reaching the impressive level of brightness achieved using narrow-band cavity QED approaches in confocal microscopy setups~\cite{Loredo2016,Ding2016}, in which the devices have fiber-coupled source efficiency in the range 13~\% to 15~\%. The planar technology promises a path towards fully-integrated photonic quantum-information processing.

In conclusion, we have presented an approach to a highly efficient single-photon source by coupling a QD to a high $\beta$-factor photonic waveguide and subsequently transferring the collected photons to an optical fiber via evanescent coupling. This is a significant step forward towards a truly deterministic source of single photons, where all  loss mechanisms associated with generation, propagation, and detection are carefully controlled. By implementing tapered photonic waveguides in conjunction with tapered fiber dimples, a chip-to-fiber coupling efficiency exceeding 80~\% was demonstrated. The overall source efficiency of the fiber-coupled device was found to be 10~\% to 15~\% and the relevant loss processes were carefully analyzed. We anticipate that, by optimizing further the chip-to-fiber coupling interface, introducing better quality microfibers \cite{Hoffman2014}, and suppressing radiative blinking and mixing of dipoles by implementing resonant excitation, a single-photon source with efficiency exceeding 50~\% is soon within reach, meaning that a rate of $>40$~MHz of single photons may be generated at the output of a single-mode fiber. Furthermore, implementing electrical control of QDs in the waveguide structures allows generation of highly coherent photons \cite{Somaschi2016,Kirsanske2016}. Such an efficient source would create new possibilities for the experimental realisation of photonic boson sampling and quantum simulation with a large number of photons \cite{He2016scalable,Loredo2016bosonsampling}. Moreover, the planar photonic-crystal platform opens a range of novel opportunities for implementing more advanced quantum functionalities for targeting the construction of complex scalable photonic networks \cite{Lodahl2016,Kimble2008quantum}.

\section*{Funding Information}
RSD, TP, LM, SS, and PL gratefully acknowledge financial support from the following funding agencies: the Lundbeck Foundation, the Villum Foundation, the Carlsberg Foundation, the European Research Council (ERC Consolidator Grant "ALLQUANTUM" and ERC Advanced Grant "SCALE"), Innovation Fund Denmark (Quantum Innovation Center "Qubiz"), and the Danish Council for Independent Research. KCB and JL acknowledge support under the Cooperative Research Agreement between the University of Maryland and NIST-CNST, Award 70NANB10H193. EHL and JDS acknowledge financial supports from 301 the KIST Internal Program of flag-ship (2E26420).

\section*{Acknowledgments}

We thank Marcelo Davan\c co for support during the experiment and Sahand Mahmoodian for fruitful discussions during the preparation of the manuscript.

\end{document}